\documentclass[letter]{aa}
\usepackage{graphicx}
\newcommand{\kmprs}  {\mbox{\rm km\,s$^{-1}$}}
\newcommand{\micro}{\mbox{$\xi_{\rm turb}$}}

\newcommand{\feh} {\mbox{\rm [Fe/H]}}

\newcommand{\nafe} {\mbox{\rm [Na/Fe]}}
\newcommand{\mgfe} {\mbox{\rm [Mg/Fe]}}
\newcommand{\sife} {\mbox{\rm [Si/Fe]}}
\newcommand{\cafe} {\mbox{\rm [Ca/Fe]}}
\newcommand{\tife} {\mbox{\rm [Ti/Fe]}}
\newcommand{\crfe} {\mbox{\rm [Cr/Fe]}}
\newcommand{\nife} {\mbox{\rm [Ni/Fe]}}

\newcommand{\alphafe} {\mbox{\rm [$\alpha$/Fe]}}
\newcommand{\fracfeh} {\mbox{${\rm [\frac{Fe}{H}]}$}}

\newcommand{\fracnafe} {\mbox{${\rm [\frac{Na}{Fe}]}$}}
\newcommand{\fracmgfe} {\mbox{${\rm [\frac{Mg}{Fe}]}$}}
\newcommand{\fracsife} {\mbox{${\rm [\frac{Si}{Fe}]}$}}
\newcommand{\fraccafe} {\mbox{${\rm [\frac{Ca}{Fe}]}$}}
\newcommand{\fractife} {\mbox{${\rm [\frac{Ti}{Fe}]}$}}
\newcommand{\fraccrfe} {\mbox{${\rm [\frac{Cr}{Fe}]}$}}
\newcommand{\fracnife} {\mbox{${\rm [\frac{Ni}{Fe}]}$}}

\newcommand{\teff}  {\mbox{$T_{\rm eff}$}}

\newcommand{\logg}  {\mbox{{\rm log}\,$g$}}

\newcommand{\MgI} {\ion{Mg}{i}}
\newcommand{\NaI} {\ion{Na}{i}}
\newcommand{\SiI} {\ion{Si}{i}}

\newcommand{\CaI} {\ion{Ca}{i}}
\newcommand{\TiI} {\ion{Ti}{i}}

\newcommand{\CrI} {\ion{Cr}{i}}

\newcommand{\FeI} {\ion{Fe}{i}}
\newcommand{\FeII} {\ion{Fe}{ii}}
\newcommand{\NiI} {\ion{Ni}{i}}

\newcommand{\Vtotal}   {\mbox{$V_{\rm total}$}}

\newcommand{\VK}{\mbox{($V\!-\!K)$}}

\newcommand{\by}{\mbox{($b\!-\!y)$}}

\def\ltsima{$\; \buildrel < \over \sim \;$}
\def\simlt{\lower.5ex\hbox{\ltsima}}
\def\gtsima{$\; \buildrel > \over \sim \;$}
\def\simgt{\lower.5ex\hbox{\gtsima}}
\begin{document}

\title{Two distinct halo populations in the solar neighborhood
\thanks{Based on observations made with the Nordic Optical Telescope
on La Palma, and on data from the European Southern Observatory
ESO/ST-ECF Science Archive Facility.}}

\subtitle{Evidence from stellar abundance ratios and kinematics}

\author{P.E.~Nissen \inst{1} \and W.J.~Schuster \inst{2}}


\institute{
Department of Physics and Astronomy, University of Aarhus, DK--8000
Aarhus C, Denmark.
\email{pen@phys.au.dk}
\and Observatorio Astron\'{o}mico Nacional, Universidad Nacional Aut\'{o}noma
de M\'{e}xico, Apartado Postal 877, C.P. 22800 Ensenada, B.C., M\'{e}xico.
\email{schuster@astrosen.unam.mx}}

\date{Received 15 December 2009 / Accepted 18 February 2010}

\abstract
{}
{Precise abundance ratios are determined for 94 dwarf stars
with $5200 < \teff < 6300$\,K, $-1.6 <$ \feh $< -0.4$, and 
distances $D \simlt 335$\,pc. Most of them have halo kinematics, but
16 thick-disk stars are included.}
{Equivalent widths of atomic lines are measured from 
VLT/UVES and NOT/FIES spectra with resolutions $R\simeq \! 55\,000$ 
and $R \simeq \! 40\,000$,
respectively. An LTE abundance analysis based on MARCS
models is applied to derive precise differential abundance ratios of
Na, Mg, Si, Ca, Ti, Cr, and Ni with respect to Fe.} 
{The halo stars fall into two populations, clearly separated in
\alphafe , where $\alpha$ refers to the average abundance of 
Mg, Si, Ca, and Ti. Differences in [Na/Fe] and [Ni/Fe] are
also present with a remarkably clear correlation between these 
two abundance ratios.} 
{The `high-$\alpha$' stars may be ancient disk or bulge stars 
`heated' to halo kinematics by merging satellite galaxies or they could have formed
as the first stars during the collapse of a proto-Galactic gas cloud. 
The kinematics of the `low-$\alpha$' stars
suggest that they have been accreted from dwarf galaxies, and that
some of them may originate from the $\omega$\,Cen progenitor galaxy.}


\maketitle

\section{Introduction}
\label{introduction} 
Studies of stellar  populations are
of high importance for understanding the formation and
evolution of the Milky Way. In this context, it has been discussed
whether the Galactic halo consists of more than one population. The
monolithic collapse model of Eggen et al.
(\cite{eggen62}) corresponds to a single 
population, but from a study of globular clusters, Searle and Zinn
(\cite{searle78}) suggested that the halo
contains two populations: $i)$ an inner, old, flattened
population with a slight prograde rotation formed during a dissipative
collapse, and $ii)$ an outer, younger, spherical population 
accreted from dwarf galaxies.
The presence of this dichotomy was supported by a study of 
$\sim \! 20\,000$ stars in the SDSS survey performed
by Carollo et al. (\cite{carollo07}).

Elemental abundances of stars in the solar neighborhood may provide 
additional information about the halo populations.
In this context, the ratio \alphafe , where $\alpha$ refers to the 
average abundance of Mg, Si, Ca, and Ti, is of particular
interest. The $\alpha$-elements are produced mainly
during Type II supernovae (SNe) explosions on a short timescale
($\sim \! 10^7$ years), whereas iron is also produced by Type Ia SNe on a
much longer timescale ($\sim \! 10^9$ years). Hence, \alphafe\ can be
used as a `clock' to probe the star formation history of
a Galactic component. 

Several previous studies have focused on the possible differences in
\alphafe\ for stars in the solar neighborhood.
Fulbright (\cite{fulbright02}), Stephens \& Boesgaard (\cite{stephens02}),
and Gratton et al. (\cite{gratton03}) all find evidence that stars
associated with the outer halo have lower
\alphafe\ than stars connected to the inner halo.
The differences in \alphafe\ found in these studies are, however,
not larger than 0.1\,dex, and it is unclear whether the distribution of \alphafe\
is continuous or bimodal. Nissen \& Schuster (\cite{nissen97}) 
achieved a higher precision measurement of \alphafe\  and
found evidence of a bimodal distribution of
\alphafe\ for 13 halo stars with $-1.3 < \feh < -0.4$, but
a larger sample of these `metal-rich' halo stars needs to be
observed to confirm these findings and study possible correlations with kinematics. 
In this Letter, we present the first results of such a study.

\section{Sample selection and observed spectra}
\label{sect:obs} 
Stars were selected from the Schuster et al.
(\cite{schuster06}) $uvby$-$\beta$ catalogue of high-velocity and metal-poor
stars. To ensure that a star has a high probability of belonging to the halo
population, the total space velocity, \Vtotal , with respect to 
the local standard of rest (LSR) was constrained to  be larger 
than 180\,\kmprs\ (Venn et al. \cite{venn04}).
Furthermore, the Str\"{o}mgren indices \by , $m_{\rm 1}$ and
$c_{\rm 1}$ were used to select dwarfs and subgiants with
$5200 < \teff < 6300$\,K and $\feh \simgt -1.6$. 
This produced a list of about 200 stars, of which
37 have VLT/UVES spectra available in the ESO/ST-ECF Science Archive
(Table \ref{table:UVES.RV}).
Furthermore, 16 stars with thick-disk kinematics and UVES spectra were
included.
In addition, 53 randomly selected stars were observed with the FIbre fed
Echelle Spectrograph (FIES) at the Nordic Optical Telescope (NOT)
(Table \ref{table:FIES.RV}).
Six were found to be double-lined spectroscopic binaries 
and excluded.
All FIES stars and most of the UVES stars are brighter than $V = 11.1$,
three having $V$ = 11.2, 12.2, and 12.8. The average distance is
115\,pc with $D_{\rm max} = 335$\,pc.

The UVES spectra cover the spectral region 4800 - 6800\,\AA\ and
have resolutions $R\simeq \! 55\,000$ and signal-to-noise ratios
($S/N$) from  250 to 500.
The FIES spectra range from 4000 to 7000\,\AA ,
but only the 4700 - 6400\,\AA\ region was employed,
with a resolution $R\simeq \! 40\,000$ and $S/N \simeq$ 140 - 200.
The majority of the UVES stars had reduced spectra
available in the archive, but for stars
observed with an image slicer, the raw data were reduced 
using the echelle package in IRAF.
The FIES data were handled by {\tt FIEStool}, a data
reduction package developed by E. Stempels.

Equivalent widths (EWs) of 130 to 180 atomic lines were measured
for each star.  The large majority of the lines have EWs
between 2 and 90\,m\AA .  For six stars, 
both UVES and FIES spectra are available. The average EW difference 
(FIES -- UVES) 
is 0.6\,m\AA\ with a rms deviation of 1.3\,m\AA .

\section{Stellar parameters and abundances}
\label{sect:par-abun}
Element abundances are derived from
EWs using the Uppsala
EQWIDH program together with model atmospheres
interpolated from the new MARCS grid (Gustafsson et
al. \cite{gustafsson08}). Two sets of models are available with different
values of \alphafe , which makes it possible to
interpolate to a model having the same \alphafe\ as the star.
Local thermodynamic equilibrium (LTE) is assumed in the line
calculations, and line broadening caused by microturbulence, \micro , and
collisional damping is included. 

The abundance analysis 
is performed differentially with respect to two bright thick-disk stars,
\object{HD\,22879} and \object{HD\,76932}. Their effective temperatures
are determined from \by\ and \VK\ using the calibrations of
Ram\'{\i}rez \& Mel\'{e}ndez (\cite{ramirez05}).
Surface gravities are derived from Hipparcos
parallaxes as described by Nissen et al. (\cite{nissen04}),
and chemical abundances 
from a differential analysis with respect to the Sun,
using a subset of $\sim \! 80$ lines, which
are relatively unblended in the solar flux spectrum
of Kurucz et al. (\cite{kurucz84}).
In an `inverted' abundance analysis, the data
from the star-Sun analysis are then used to determine
$gf$-values for the whole set of  $\sim \! 180$ lines.
These $gf$-values are applied for the abundance analysis 
of all program stars. 

We then determine \teff\ so that the
\feh\ derived from the \FeI\ lines is independent of 
excitation potential.  As the \FeI\ lines are also
used to determine \micro\ by minimizing the dependence of \feh\  
on EW, iteration is needed
to obtain consistent values of \teff\ and \micro . 
We estimate a differential error of $\sigma (\teff )$ = $\pm 30$\,K
by comparing \teff\ values derived from the \FeI\ excitation balance
with those inferred from \by\ and \VK\ colors 
for a subset of 44 nearby stars that appear to be
unreddened according to the absence of interstellar NaD lines. 
The surface gravity is estimated by ensuring that
\FeI\ and \FeII\ lines provide consistent Fe abundances.
Comparison of these spectroscopic gravities
with values determined from Hipparcos parallaxes
for the nearby stars shows that \logg\ is determined differentially
to a precision of 0.05\,dex. 

The derived abundance ratios of Na, Mg, Si, Ca, Ti, Cr, and Ni with
respect to Fe are given in Tables \ref{table:UVES} and \ref{table:FIES}.
All abundance ratios are based on neutral lines.
The numbers of the lines applied are
\NaI\ 2-5, \MgI\ 1-2, \SiI\ 5-10, \CaI\ 6-9, \TiI\ 9-14, \CrI\ 4-7, \FeI\ 70-92, \FeII\ 14-16, and \NiI\ 20-27, where the first number refers to 
the most metal-poor stars, and the last to the most metal-rich.
 
The errors in the abundance
ratios were estimated by comparing results obtained from
UVES and FIES spectra for the six stars observed with
both instruments (see Tables \ref{table:UVES} and \ref{table:FIES}).
The spectra of these stars have typical $S/N$,
except \object{HD\,189558} that has an unusually high quality FIES spectrum
($S/N \simeq 350$).
This comparison  shows that differential values of \feh , \nafe , \mgfe , and \sife\
are determined to a 1-$\sigma$ precision of 0.03 to 0.04\,dex, whereas 
the precision of \cafe , \tife , \crfe , and \alphafe\ is about 0.02\,dex.
The error in \nife\ is as small as 0.01\,dex, because of the many
\FeI\ and \NiI\ lines available. We note that errors in
the abundance ratios caused by errors in \teff\ and \logg\ are
small compared to errors induced by the EW measurements, because all ratios
are derived from neutral atomic lines with similar sensitivity to \teff\ and \logg .

\begin{figure}
\resizebox{\hsize}{!}{\includegraphics{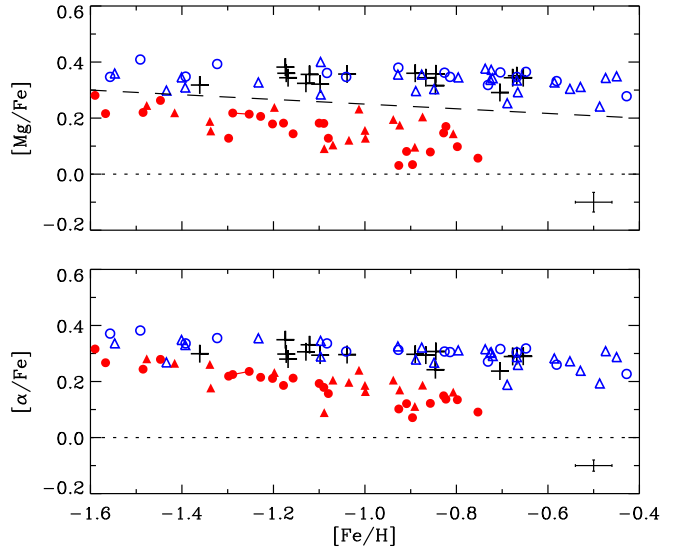}}
\caption{\mgfe\ and \alphafe\ versus \feh . Crosses refer to thick-disk stars
and circles to halo stars observed with UVES. Triangles indicate halo
stars with FIES spectra. Halo stars above the long-dashed line in the 
\mgfe\ diagram are defined as belonging to the high-$\alpha$  population
and are indicated by open (blue) symbols. 
The stars below the long-dashed line are defined to be low-$\alpha$ 
stars and are shown with filled (red) symbols.
Based on \mgfe , this classification is maintained in all the following figures.
The components of a visual binary star, 
\object{G\,112-43} and \object{G\,112-44}, are
connected by a straight line.} 
\label{fig:mg.alpha-fe}
\end{figure}

Figure \ref{fig:mg.alpha-fe} shows \mgfe\ and \alphafe\
as a function of \feh . We note that there are no systematic offsets
between the UVES and the FIES data. The corresponding figure for
\sife , \cafe , and \tife\ is shown in the Online Section.
As can be seen, the halo stars consist of two distinct populations,
the `high-$\alpha$' stars with a nearly constant \alphafe\
and the `low-$\alpha$' stars with a declining
\alphafe\ as a function of increasing metallicity. A
classification into these two populations was performed
on the basis of \mgfe . In the range $-1.6 < \feh
< -1.4$, the two populations tend to merge, and the classification
is less clear. The high-$\alpha$ and low-$\alpha$ halo populations 
also separate well in \nafe\ and \nife\ with the exceptions of two Na-rich stars.
The abundance differences can be seen directly from the observed 
spectra as shown in the Online Section.

\begin{figure}
\resizebox{\hsize}{!}{\includegraphics{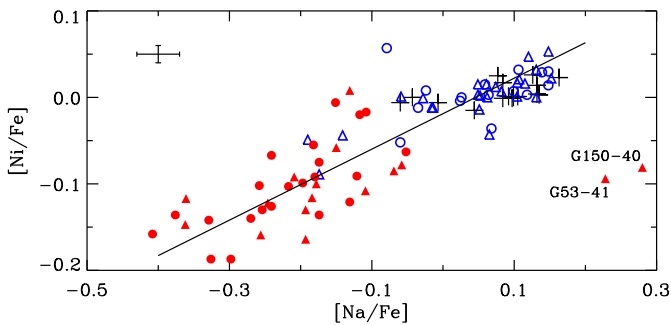}}
\caption{\nife\ versus \nafe\ with the same symbols
as in Fig. \ref{fig:mg.alpha-fe}. The linear fit
does not include the two Na-rich stars.}
\label{fig:ni-na}
\end{figure}

The scatter in the abundance ratios for the high-$\alpha$
and thick-disk stars relative to the best-fit linear relations
is 0.032\,dex for \mgfe\ and 0.030\,dex for \alphafe . 
This is similar to the estimated errors of the analysis.
For the low-$\alpha$ stars, there are, however, abundance differences
from the trends that cannot be explained by the errors alone,
especially in the case of  \nafe\ and \nife . The clear correlation
between these ratios (Fig.  \ref{fig:ni-na}) confirms that
cosmic variations in these ratios are present at a given \feh .

\section{Kinematics}
\label{sect:kinematics}
To calculate the stellar space velocities, we acquired
proper motions from
the Tycho--2 catalogue (H{\o}g et al. \cite{hoeg00}, 88 stars),
the new reduction of the Hipparcos data (van Leeuwen \cite{leeuwen07}, 4 stars),
and the revised NLTT (Salim \& Gould \cite{salim03}, 2 stars).
Distances of stars were calculated from the parallaxes
of van Leeuwen (\cite{leeuwen07}), when the errors in these are less than
10\%, and if not, from the photometric absolute magnitude calibration
by Schuster et al.  (\cite{schuster04}, \cite{schuster06}). 
The radial velocities of the stars were derived from 
our own spectra and have errors of $\pm 0.3$\,\kmprs .

With these data as input, the formulae and matrix equations of 
Johnson \& Soderblom (\cite{johnson87}) were used to calculate
the Galactic velocity components ($U, V, W$) and their errors. 
Correction for a solar motion of (+7.5, +13.5, +6.8)\,\kmprs\ 
with respect to the LSR was adopted from
Francis \& Anderson (\cite{francis09}). The resulting values of
$U_{\rm LSR}, V_{\rm LSR}$, and $W_{\rm LSR}$ are given in 
Tables \ref{table:UVES} and \ref{table:FIES}.
The average errors of these velocities for the halo stars
are $(\pm 12, \pm 16, \pm 9)$\,\kmprs\ with a major contribution
from the error in the distances.

\begin{figure}
\resizebox{\hsize}{!}{\includegraphics{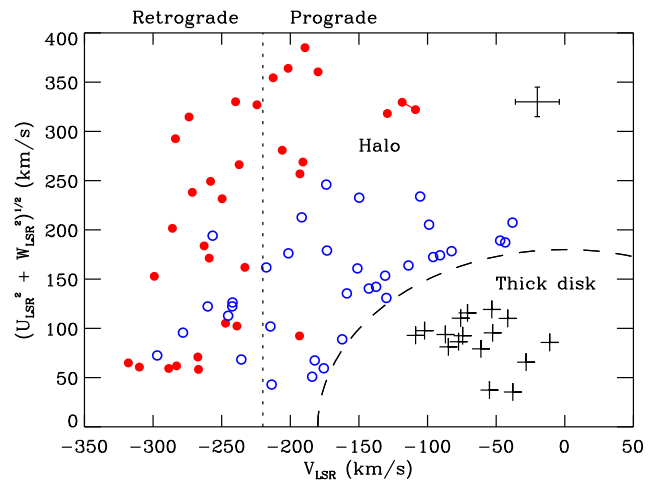}}
\caption{Toomre diagram for stars with $\feh > -1.4$.
{\rm High-$\alpha$ halo stars are shown with open (blue) circles, 
low-$\alpha$ halo stars with filled (red) circles, and thick-disk stars
with crosses.}  The long-dashed line corresponds to $V_{\rm total} = 180$\,\kmprs .
The short-dashed line indicates zero rotation in the Galaxy.}
\label{fig:Toomre}
\end{figure}

Figure \ref{fig:Toomre} shows the Toomre diagram for the thick-disk and
halo stars that could be clearly classified as belonging to
either the high-$\alpha$ or the low-$\alpha$ population. 
Assuming Gaussian velocity
distributions with canonical dispersions and asymmetric drifts 
for the thin-disk, thick-disk, and halo populations, stars with
$\Vtotal > 180$\,\kmprs\ generally have a high probability of belonging
to the halo population (Venn et al. \cite{venn04}).
If, on the other hand, the velocity distribution of the thick disk is
non-Gaussian with an extended tail toward high velocities, as in
the model of the Galactic disks by Sch\"{o}nrich \& Binney
(\cite{schonrich09}), then the high-$\alpha$ stars with 
$180 < \Vtotal < 210$\,\kmprs\ might belong to the thick-disk population.
Nevertheless, the remaining high-$\alpha$ halo stars exhibit a
well-defined trend that is clearly separated from that of 
the low-$\alpha$ stars.

\section{Discussion}
\label{sect:discussion}
As discussed in detail by Gilmore \& Wyse
(\cite{gilmore98}), the near-constancy of \alphafe\ 
for the high-$\alpha$ and thick-disk stars suggests that they
formed in regions with such a high star formation rate that only
Type II SNe contributed to their chemical enrichment up to
$\feh \simeq -0.4$.
On the other hand, the low-$\alpha$ stars originate in
regions with a relatively slow chemical evolution so that
Type Ia SNe have started to contribute iron at
$\feh \simeq -1.5$ causing \alphafe\ to decrease toward higher
metallicities.

The distinction between the two halo populations is 
greater for \mgfe\ than for both \cafe\ and \tife , 
probably because of different SNe Ia yields.
According to Tsujimoto et al. (\cite{tsujimoto95}),
the relative contributions of SNe Ia to the solar abundances
are negligible for Mg, 17\% for Si, 25\% for Ca, and
57\% for Fe. 

As discussed by Venn et al. (\cite{venn04}), the yields of the
neutron-rich isotopes $^{23}$Na and $^{58}$Ni from massive stars
is controlled by the neutron excess, which depends on the initial 
heavy-element abundance (Arnett \cite{arnett71}). 
It would be interesting to investigate
in more detail whether these dependences could explain
the underabundances of Na and Ni in low-$\alpha$ stars
and the correlation seen in Fig. \ref{fig:ni-na}.

As seen in the Toomre diagram, the high-$\alpha$ stars show
evidence of being more bound to the Galaxy and
favoring prograde Galactic orbits, while the low-$\alpha$ 
stars are less bound with two-thirds of them being on
retrograde orbits. This suggests that the high-$\alpha$
population is connected to a dissipative component of the Galaxy
that experienced rapid chemical evolution similar to that of the thick disk, 
whereas the low-$\alpha$ stars were accreted
from dwarf galaxies that had lower star formation rates.

Present-day dwarf spheroidal galaxies tend to have even
lower values of \alphafe , \nafe , and \nife\  
than the low-$\alpha$ halo stars for the range $-1.6 < \feh < -0.8$
(Tolstoy et al. \cite{tolstoy09}). This offset agrees with 
the predictions of the simulations of
a hierarchically formed stellar halo in a
$\Lambda$CDM Universe by Font et al. (\cite{font06}). 
The bulk of halo stars originate from early accreted,
massive dwarf galaxies with efficient star formation, whereas 
surviving satellite galaxies in the outer halo are on average 
of lower mass and experience slower chemical evolution with a greater
contribution from Type Ia SNe at a given metallicity.
The predicted \mgfe\ versus \feh\ relation for the accreted halo stars agrees
remarkably well with the trend for the low-$\alpha$ halo stars. However,
Font et al. do not explain the existence of high-$\alpha$ halo stars.
Two $\Lambda$CDM simulations suggest a dual origin
of stars in the inner Galactic halo. Purcell et al. (\cite{purcell09})
propose that ancient stars formed in the Galactic disk may be ejected
into the halo by the merging of satellite galaxies, and
Zolotov et al. (\cite{zolotov09}) find that stars formed out of accreted
gas in the inner 1\,kpc of the Galaxy can be displaced into the halo through
a succession of mergers. Alternatively, the high-$\alpha$ population
might have formed as the first stars in a dissipative collapse
of a proto-Galactic gas cloud (Gilmore et al. \cite{gilmore89};
Schuster et al. \cite{schuster06}, Sect. 8.2).

\begin{figure}
\resizebox{\hsize}{!}{\includegraphics{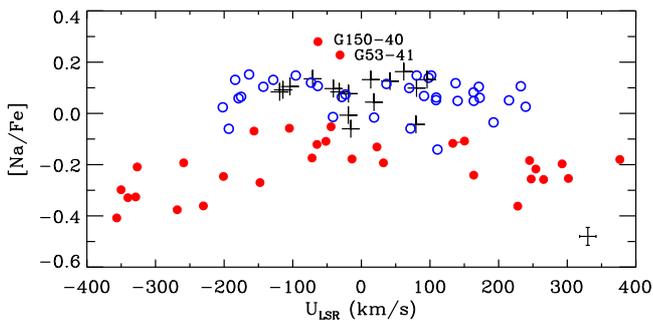}}
\caption{\nafe\ versus $U_{\rm LSR}$ for stars with $\feh > -1.4$.
The same symbols as in Fig. \ref{fig:Toomre} are used.}.
\label{fig:na-U}
\end{figure}

The retrograde low-$\alpha$ stars in Fig. \ref{fig:Toomre}
have an average Galactic rotation velocity of 
$V_{\rm LSR} \simeq -260$\,\kmprs , which is close to that of the
$\omega$\,Cen globular cluster (Dinescu et al. \cite{dinescu99}).
As often discussed (e.g., Bekki \& Freeman \cite{bekki03}), $\omega$\,Cen is
probably the nucleus of a captured satellite galaxy with its
own chemical enrichment history. Meza et al. (\cite{meza05}) 
simulated the orbital characteristics of the tidal debris of
such a satellite dragged into the
Galactic plane by dynamical friction. The captured stars
have rather small $W$-velocities but a wide, double-peaked
$U$-distribution, similar to the $W$-$U$  distribution
observed for the low-$\alpha$ halo (see Online Section). 
As shown in Fig. \ref{fig:na-U}, stars with
extreme $U$ velocities tend to have the lowest \nafe\ values,
which corroborates their having a special origin.

In support of a connection between low-$\alpha$ stars and
$\omega$\,Cen, we note that stars in this globular cluster
exhibit a wide range of \feh\ values and a decline in \alphafe\
for metallicities above $\feh \sim -1$ (Origlia et al. \cite{origlia03}). 
Johnson et al.  (\cite{johnson09}), on the other hand,
find that  \nafe\ in $\omega$\,Cen red giants increases from about  $-0.2$\,dex
at $\feh \sim -1.7$ to +0.8\,dex at $\feh \sim -1.0$.
A similar increase is not seen for the low-$\alpha$ halo stars. 
Enhancements of Na and a
Na-O anticorrelation are present in all well-studied globular clusters
(Carretta et al. \cite{carretta09}) and may be caused by the
chemical enrichment from intermediate-mass AGB stars undergoing
hot-bottom hydrogen burning. According to the hydrodynamical
simulations of D'Ercole et al. (\cite{dercole08}), the gas ejected
from these AGB stars collects in the cluster core via cooling flows,
which may explain the difference in \nafe\ between stars remaining in 
$\omega$\,Cen itself and those originating in the progenitor galaxy. 

We conclude that the derived abundance ratios provide clear evidence
of two distinct populations of stars that are among the most 
metal-rich in the Galactic halo. The reason that previous studies
have failed to detect this dichotomy may be ascribed to the lower precision
of the abundances for less homogeneous samples of stars,
and greater focus on metal-poor stars. 
The high-$\alpha$
stars may be ancient disk or bulge stars `heated' to halo kinematics
by merging satellite galaxies or they could be the first stars 
formed in a dissipative
collapse of a proto-Galactic gas cloud. The low-$\alpha$ stars are probably
accreted from dwarf galaxies, and some are likely
to be associated with the $\omega$\,Cen progenitor galaxy.
Further studies of possible
correlations between the abundance ratios and orbital parameters of
the stars may help us to clarify the origin of the two populations.

\begin{acknowledgements}
We thank the anonymous referee for comments and suggestions, which
helped to improve this Letter significantly.
\end{acknowledgements}

\Online

\begin{figure}
\resizebox{\hsize}{!}{\includegraphics{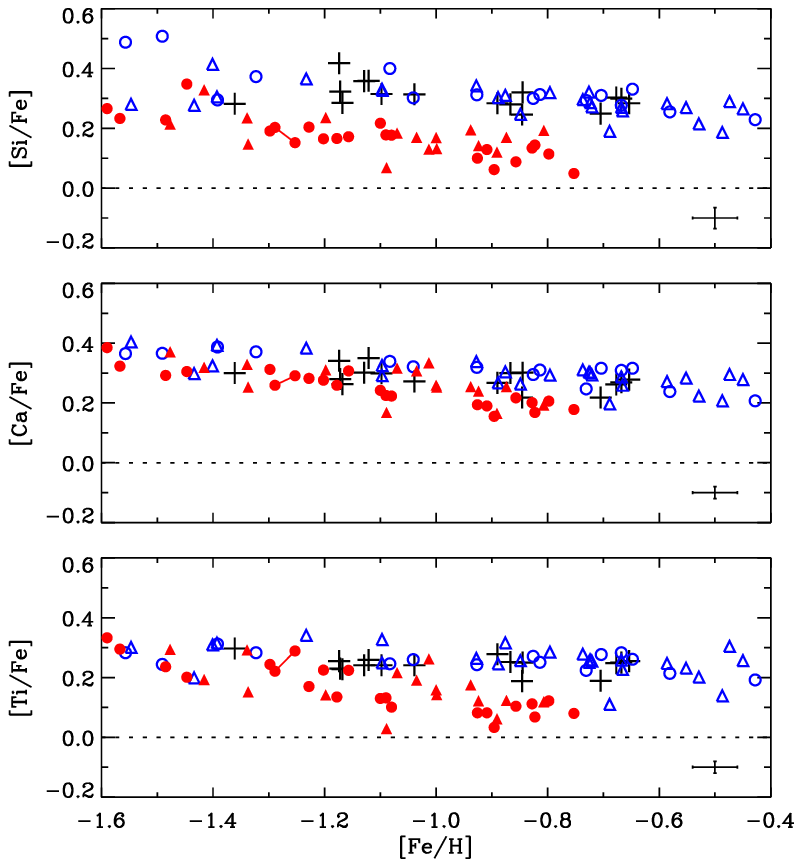}}
\caption{\sife , \cafe , and \tife\ as a function of \feh .
The same symbols as in Fig. \ref{fig:mg.alpha-fe} are used.}
\label{fig:si.ca.ti-fe}
\end{figure}

\begin{figure}
\resizebox{\hsize}{!}{\includegraphics{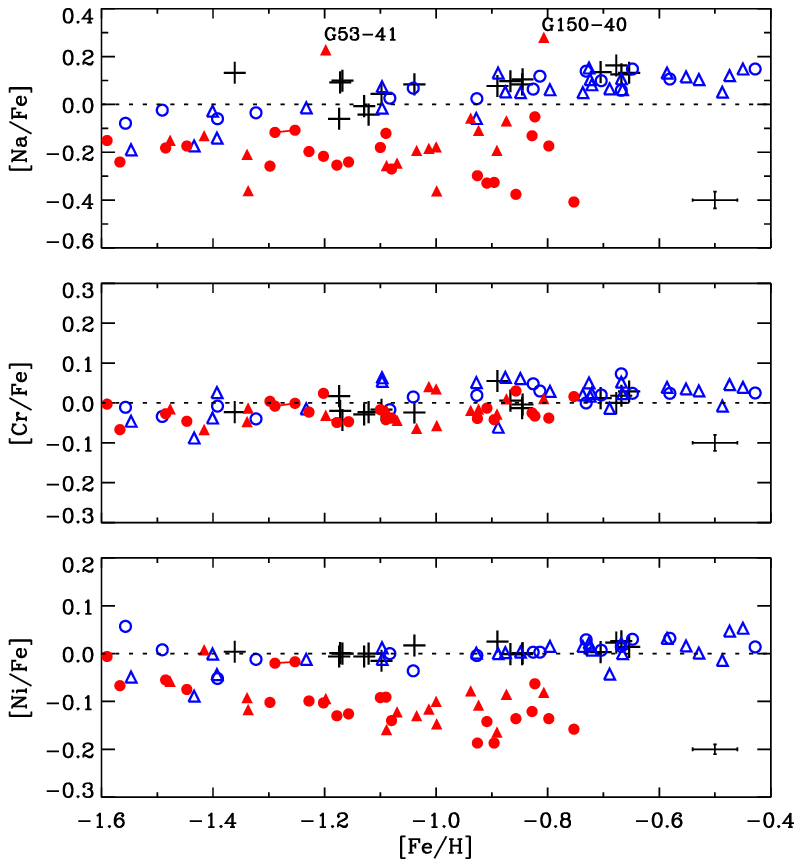}}
\caption{\nafe , \crfe , and \nife\ as a function of \feh .
The same symbols as in Fig. \ref{fig:mg.alpha-fe} are used.}
\label{fig:na.cr.ni-fe}
\end{figure}

\begin{figure}
\resizebox{\hsize}{!}{\includegraphics{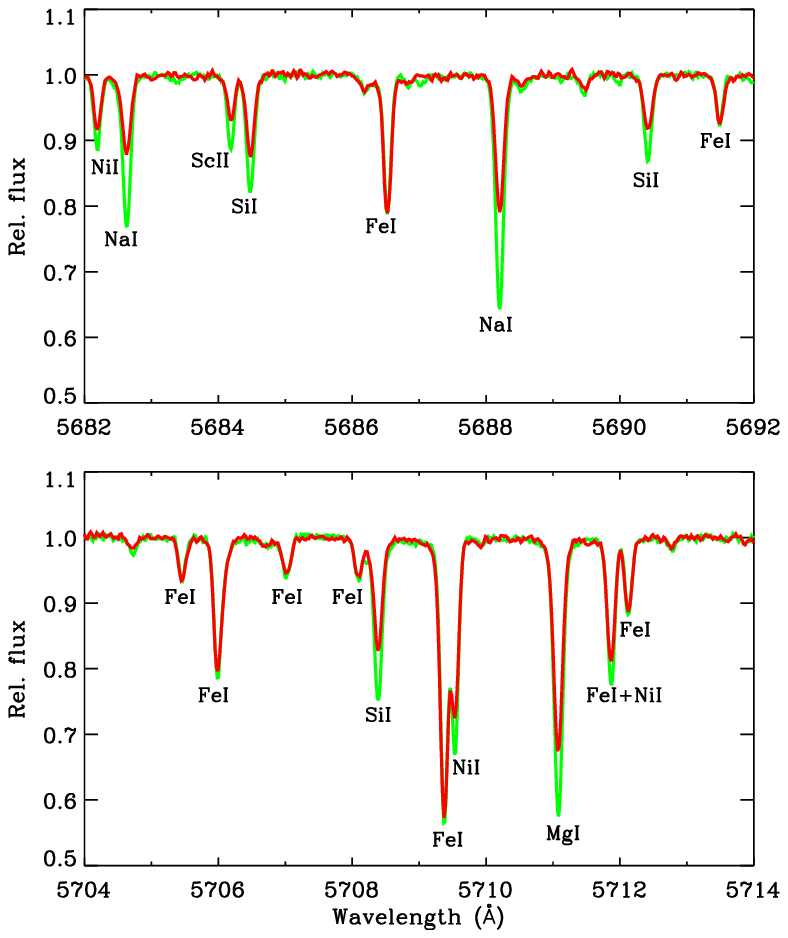}}
 \caption{UVES spectra of two stars with nearly the same atmospheric parameters
\teff , \logg , and \feh .  The spectrum of the low-$\alpha$ star
\object{CD\,$-45\,3283$} (\teff \,=\,5597\,K, \logg \,=\,4.55, \feh
\,=\,$-0.91$, \alphafe \,=\,0.12) is shown with a red line
and that of the high-$\alpha$ star
\object{G\,159-50} (\teff \,=\,5624\,K, \logg \,=\,4.37, \feh
\,=\,$-0.93$, \alphafe \,=\,0.31) with a green line. The Fe lines
have the same strength in the two spectra, but the Na, Mg, Si, and Ni
lines are significantly weaker in the spectrum of the low-$\alpha$ star.}
\label{fig:UVES.spectra}
\end{figure}

\begin{figure}
\resizebox{\hsize}{!}{\includegraphics{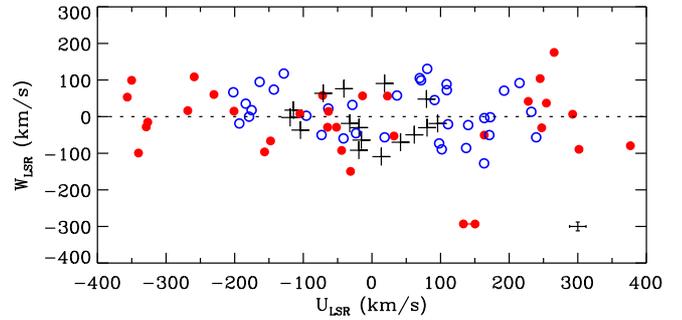}}
\caption{$W_{\rm LSR}$ versus $U_{\rm LSR}$ for stars with $\feh > -1.4$.
The same symbols as in Fig. \ref{fig:Toomre} are used.}
\label{fig:W-U}
\end{figure}

\begin{table*}
\caption[ ]{UVES spectra acquired from the ESO/ST-ECF Science Archive.}
\label{table:UVES.RV}
\setlength{\tabcolsep}{0.15cm}
\begin{tabular}{lcccrccc}
\noalign{\smallskip}
\hline\hline
\noalign{\smallskip}
  ID & Program  & Date and  UT  & $S/N$  & RV$^{\rm a}$ & $EW(\NaI\,$D$_2)^{\rm b}$ & $EW(\NaI\,$D$_1)^{\rm b}$ & Note$^{\rm c}$ \\
     &          &               &        &(\kmprs )& (m\AA ) & (m\AA )&    \\
\hline
\noalign{\smallskip}
BD$-$21 3420 &   76.B-0133 &  2006-01-14T08:28 &   300 & $   7.9$ & $ <3 $ & $ <3 $ &    \\
CD$-$33 3337 &   68.D-0094 &  2001-11-26T07:43 &   600 & $  73.6$ & $ <3 $ & $ <3 $ &    \\
CD$-$43 6810 &   70.D-0474 &  2002-12-07T07:58 &   330 & $ 159.2$ & $ 62 $ & $ 53 $ &    \\
CD$-$45 3283 &   69.D-0679 &  2002-02-19T01:53 &   300 & $ 309.2$ & $ <3 $ & $  ? $ &    \\
CD$-$51 4628 &   69.D-0679 &  2002-02-18T06:15 &   280 & $ 197.9$ & $ 15 $ & $  6 $ &    \\
CD$-$57 1633 &   69.D-0679 &  2002-02-19T01:27 &   450 & $ 260.7$ & $ <3 $ & $  ? $ &    \\
CD$-$61 282  &   76.B-0133 &  2005-09-22T06:35 &   280 & $ 220.4$ & $  ? $ & $ 10:$ &    \\
G05-19     &   76.B-0133 &  2005-10-15T05:21 &   300 & $-216.2$ & $150 $ & $130 $ &    \\
G05-40     &   68.B-0475 &  2001-10-08T06:25 &   450 & $-117.6$ & $196 $ & $165 $ &    \\
G18-28     &   76.B-0133 &  2005-10-13T00:13 &   300 & $-203.1$ & $ <3 $ & $ <3 $ &  SB1 \\
G18-39     &   71.B-0529 &  2003-08-19T07:26 &   480 & $-234.1$ & $ 69 $ & $ 38 $ &    \\
G20-15     &   71.B-0529 &  2003-08-11T03:08 &   500 & $  84.3$ & $198 $ & $172 $ &    \\
G46-31     &   76.B-0133 &  2005-12-24T03:59 &   300 & $ 220.9$ & $ <3 $ & $ <3 $ &  SB1 \\
G63-26     &   72.B-0585 &  2004-03-23T06:52 &   240 & $  57.4$ & $ 29+19 $ & $ 19+6 $ & \\
G66-22     &   69.D-0679 &  2002-02-18T08:57 &   250 & $-145.2$ & $ 12 $ & $  6 $ &    \\
G82-05     &   69.D-0679 &  2002-02-16T01:22 &   300 & $ 296.0$ & $ <3 $ & $ <3 $ &    \\
G112-43    &   69.D-0679 &  2002-02-17T02:13 &   340 & $ -83.7$ & $ 16 $ & $  ? $ &    \\
G112-44    &   70.D-0474 &  2002-12-31T04:56 &   350 & $ -84.3$ & $ 14 $ & $  7 $ &    \\
G114-42    &   69.D-0679 &  2002-02-17T02:58 &   250 & $ -86.4$ & $ 64 $ & $ 37 $ &    \\
G121-12    &   76.B-0133 &  2006-01-14T07:14 &   300 & $ 197.4$ & $ <3 $ & $ <3 $ &    \\
G188-22    &   71.B-0529 &  2003-08-09T06:08 &   450 & $ -93.9$ & $ 43 $ & $ 18 $ &    \\
G159-50    &   76.B-0133 &  2005-10-14T04:17 &   340 & $  28.0$ & $ <3 $ & $ <3 $ &    \\
HD3567     &   68.B-0475 &  2001-10-08T04:19 &   650 & $ -47.8$ & $ <3 $ & $ <3 $ &    \\
HD17820    &   76.B-0133 &  2005-10-13T06:39 &   400 & $   6.3$ & $ <3 $ & $ <3 $ &    \\
HD22879    &   68.D-0094 &  2001-11-26T03:27 &   450 & $ 121.0$ & $ <3 $ & $ <3 $ &    \\
HD25704    &   68.D-0094 &  2001-11-28T08:24 &   500 & $  56.0$ & $ <3 $ & $ <3 $ &    \\
HD51754    &   76.B-0133 &  2005-11-10T08:23 &   300 & $ -94.0$ & $ <3 $ & $ <3 $ &    \\
HD59392    &   67.D-0086 &  2001-03-07T03:53 &   410 & $ 268.4$ & $ <3 $ & $ <3 $ &    \\
HD76932    &   67.D-0439 &  2001-04-10T23:35 &   600 & $ 119.9$ & $ <3 $ & $ <3 $ &    \\
HD97320    &   65.L-0507 &  2000-04-09T02:09 &   460 & $  53.5$ & $ <3 $ & $ <3 $ &    \\
HD103723   &   65.L-0507 &  2000-04-10T02:11 &   500 & $ 168.3$ & $ 32 $ & $ 18:$ &    \\
HD105004   &   68.B-0475 &  2002-01-12T07:32 &   500 & $ 121.8$ & $ 40 $ & $ 21 $ &    \\
HD106516   &   76.B-0133 &  2006-01-30T07:04 &   350 & $  10.6$ & $ <3 $ & $ <3 $ &  SB1  \\
HD111980   &   65.L-0507 &  2000-04-10T03:12 &   550 & $ 159.0$ & $ 16 $ & $  ? $ &  SB1  \\
HD113679   &   65.L-0507 &  2000-04-10T04:39 &   500 & $ 158.0$ & $ 42:$ & $  ? $ &    \\
HD114762A  &   77.B-0507 &  2006-06-08T01:09 &   475 & $  49.1$ & $ <3 $ & $ <3 $ &  SB1  \\
HD120559   &   68.B-0475 &  2002-02-21T08:17 &   600 & $  18.6$ & $ <3 $ & $ <3 $ &    \\
HD121004   &   67.D-0439 &  2001-04-11T05:00 &   500 & $ 244.8$ & $ <3 $ & $ <3 $ &    \\
HD126681   &   65.L-0507 &  2000-04-09T06:31 &   500 & $ -45.3$ & $ <3 $ & $ <3 $ &    \\
HD132475   &   65.L-0507 &  2000-04-12T07:00 &   500 & $ 176.6$ & $ 28 $ & $ 16 $ &    \\
HD148816   &   77.B-0507 &  2006-06-15T03:07 &   400 & $ -47.8$ & $ <3 $ & $ <3 $ &    \\
HD163810   &   69.D-0679 &  2002-04-27T08:23 &   300 & $ 185.9$ & $ 81 $ & $ 72 $ &    \\
HD175179   &   65.L-0507 &  2000-04-10T08:46 &   400 & $  21.9$ & $ <3 $ & $ <3 $ &    \\
HD179626   &   77.B-0507 &  2006-06-12T06:01 &   330 & $ -65.0$ & $ 43 $ & $ 24 $ &    \\
HD189558   &   65.L-0507 &  2000-04-09T09:04 &   600 & $ -12.4$ & $ <3 $ & $ <3 $ &    \\
HD193901   &   77.B-0507 &  2006-06-12T06:21 &   380 & $-171.2$ & $ <3 $ & $ <3 $ &    \\
HD194598   &   77.B-0507 &  2006-06-25T03:33 &   270 & $-247.0$ & $ <3 $ & $ <3 $ &    \\
HD199289   &   77.B-0507 &  2006-06-04T01:25 &   290 & $  -5.6$ & $ <3 $ & $ <3 $ &    \\
HD205650   &   65.L-0507 &  2000-04-12T09:45 &   480 & $-105.7$ & $ <3 $ & $ <3 $ &    \\
HD219617   &   77.B-0507 &  2006-05-29T08:46 &   370 & $  13.9$ & $ <3 $ & $ <3 $ &    \\
HD222766   &   76.B-0133 &  2005-10-13T00:44 &   310 & $ -85.5$ & $ <3 $ & $ <3 $ &    \\
HD241253   &   76.B-0133 &  2005-10-08T07:52 &   300 & $ -15.7$ & $ <3 $ & $ <3 $ &    \\
HD284248   &   70.D-0474 &  2002-12-31T02:50 &   500 & $ 338.9$ & $ 71 $ & $ 63 $ &    \\
\noalign{\smallskip}
\hline
\end{tabular}

\begin{list}{}{}
\item[$^{\rm a}$]
Heliocentric radial velocity.
\end{list}
\begin{list}{}{}
\item[$^{\rm b}$]
Equivalent widths of interstellar NaD lines. A question mark means that the line is
so badly blended by a stellar or telluric line that it could not be measured.
\end{list}
\begin{list}{}{}
\item[$^{\rm c}$]
Single-lined spectroscopic binaries according to the SIMBAD database 
are designated with SB1.
\end{list}

\end{table*}

\begin{table*}
\caption[ ]{Stars observed with the NOT/FIES spectrograph.}
\label{table:FIES.RV}
\setlength{\tabcolsep}{0.15cm}
\begin{tabular}{lcccrccc}
\noalign{\smallskip}
\hline\hline
\noalign{\smallskip}
  ID & Program  & Date and  UT  & $S/N$  & RV$^{\rm a}$ & $EW(\NaI\,$D$_2)^{\rm b}$ & $EW(\NaI\,$D$_1)^{\rm b}$ & Note$^{\rm c}$ \\
     &          &               &        &(\kmprs )&  (m\AA ) & (m\AA )  &    \\
\hline
\noalign{\smallskip}
G05-36     &   38-013 &  2008-11-30T01:53 &   170 & $  -9.6$ & $189 $ & $164 $ &    \\
G13-38     &   37-003 &  2008-05-14T23:55 &   160 & $ 153.4$ & $ <5 $ & $ <5 $ &    \\
G15-23     &   37-003 &  2008-05-18T23:59 &   180 & $ -63.6$ & $ 53 $ & $ 40 $ &    \\
G16-20     &   37-003 &  2008-05-15T01:55 &   185 & $ 170.8$ & $131 $ & $ 95 $ &    \\
G20-15$^{\rm d}$ &   37-003 &  2008-05-21T01:39 &   170 & $  84.3$ & $206 $ & $172 $ &    \\
G21-22     &   37-003 &  2008-05-16T19:24 &   160 & $  59.9$ & $ 94 $ & $ 69 $ &    \\
G24-13     &   37-003 &  2008-05-17T04:50 &   170 & $  99.3$ & $110 $ & $ 89 $ &    \\
G24-25     &   37-003 &  2008-05-21T03:12 &   160 & $-312.9$ & $  ? $ & $166 $ & SB1   \\
G31-55     &   38-013 &  2008-12-04T20:56 &   160 & $ -27.9$ & $ 30 $ & $ 20 $ &    \\
G49-19     &   38-013 &  2008-11-30T14:06 &   180 & $  75.5$ & $ 63 $ & $ 46 $ & SB1    \\
G53-41     &   38-013 &  2008-12-04T05:41 &   180 & $  88.4$ & $ 48 $ & $ 28 $ &    \\
G56-30     &   37-003 &  2008-05-20T23:48 &   185 & $  27.9$ & $ <5 $ & $ <5 $ &    \\
G56-36     &   37-003 &  2008-05-18T21:16 &   175 & $  99.0$ & $ 23 $ & $ 17 $ &    \\
G57-07     &   37-003 &  2008-05-14T22:27 &   170 & $  28.6$ & $ 49 $ & $ 36 $ &    \\
G74-32     &   38-013 &  2008-12-04T11:54 &   140 & $   3.2$ & $  ? $ & $  ? $ &    \\
G75-31     &   38-013 &  2008-11-29T23:56 &   170 & $  57.7$ & $ 26 $ & $ 14 $ &    \\
G81-02     &   38-013 &  2008-12-04T14:21 &   140 & $  87.9$ & $120 $ & $ 92 $ &    \\
G85-13     &   38-013 &  2008-12-02T02:36 &   160 & $ 174.3$ & $ <5 $ & $ <5 $ &    \\
G87-13     &   38-013 &  2008-12-02T04:33 &   180 & $ 206.6$ & $105+94 $ & $67+48$ &    \\
G94-49     &   38-013 &  2008-12-02T00:30 &   170 & $-163.9$ & $ <5 $ & $ <5 $ &    \\
G96-20     &   38-013 &  2008-12-01T00:39 &   180 & $ 105.7$ & $ 28:$ & $ 12 $ &    \\
G98-53     &   38-013 &  2008-12-04T16:18 &   160 & $ 144.7$ & $136 $ & $102 $ &    \\
G99-21     &   38-013 &  2008-12-03T19:42 &   140 & $ 128.4$ & $ <5 $ & $ <5 $ &    \\
G119-64    &   37-003 &  2008-05-17T21:08 &   210 & $-195.6$ & $ <5 $ & $ <5 $ &    \\
G125-13    &   37-003 &  2008-05-19T02:49 &   175 & $-175.9$ & $ 28 $ & $ 18 $ &    \\
G127-26    &   38-013 &  2008-11-29T19:59 &   185 & $ -45.7$ & $116 $ & $ 86 $ &    \\
G150-40    &   37-003 &  2008-05-17T08:10 &   170 & $ -50.5$ & $ 31 $ & $ 14 $ &    \\
G161-73    &   37-003 &  2008-05-20T21:56 &   155 & $ 121.2$ & $134 $ & $127 $ &    \\
G170-56    &   37-003 &  2008-05-18T01:42 &   180 & $-239.9$ & $ 47 $ & $ 27 $ &    \\
G172-61    &   38-013 &  2008-11-30T21:43 &   170 & $-206.6$ & $ <5 $ & $ <5 $ & SB1   \\
G176-53    &   37-003 &  2008-05-17T22:24 &   170 & $  64.9$ & $ <5 $ & $ <5 $ &    \\
G180-24    &   37-003 &  2008-05-18T01:34 &   170 & $-152.0$ & $ <5 $ & $ <5 $ &    \\
G187-18    &   38-013 &  2008-12-01T19:15 &   170 & $-122.4$ & $ 15 $ & $  8 $ &    \\
G192-43    &   38-013 &  2008-11-30T04:04 &   180 & $ 191.2$ & $ 49 $ & $ 35 $ &    \\
G232-18    &   37-003 &  2008-05-20T04:50 &   200 & $-261.2$ & $ <5 $ & $ <5 $ &    \\
HD148816$^{\rm d}$ &   37-003 &  2008-05-16T01:33 &   200 & $ -47.3$ & $ <5 $ & $ <5 $ &    \\
HD159482   &   37-003 &  2008-05-16T02:02 &   150 & $-138.4$ & $ <5 $ & $ <5 $ &    \\
HD160693   &   37-003 &  2008-05-15T05:11 &   160 & $  34.5$ & $  ? $ & $  ? $ &    \\
HD177095   &   37-003 &  2008-05-19T04:04 &   160 & $  90.8$ & $ <5 $ & $ <5 $ &    \\
HD179626$^{\rm d}$ &   37-003 &  2008-05-20T03:14 &   170 & $ -64.9$ & $ 46 $ & $ 27 $ &    \\
HD189558$^{\rm d}$ &   37-003 &  2008-05-21T05:14 &   350 & $ -12.6$ & $ <5 $ & $ <5 $ &    \\
HD193901$^{\rm d}$ &   37-003 &  2008-05-19T05:20 &   200 & $-171.2$ & $ <5 $ & $ <5 $ &    \\
HD194598$^{\rm d}$ &   37-003 &  2008-05-15T05:33 &   180 & $-246.8$ & $ <5 $ & $ <5 $ &    \\
HD230409   &   37-003 &  2008-05-15T03:54 &   165 & $  -1.9$ & $ <5 $ & $ <5 $ &    \\
HD233511   &   38-013 &  2008-12-01T21:48 &   180 & $  66.2$ & $ <5 $ & $ <5 $ &    \\
HD237822   &   37-003 &  2008-05-18T22:20 &   175 & $  -2.3$ & $ <5 $ & $ <5 $ &    \\
HD250792A  &   38-013 &  2008-12-01T02:17 &   160 & $-191.3$ & $ <5 $ & $ <5 $ &    \\
\noalign{\smallskip}
\hline
\end{tabular}

\begin{list}{}{}
\item[$^{\rm a}$]
Heliocentric radial velocity.
\end{list}

\begin{list}{}{}
\item[$^{\rm b}$]
Equivalent widths of interstellar NaD lines. A question mark means that the line is
so badly blended by a stellar or telluric line that it could not be measured.
\end{list}

\begin{list}{}{}
\item[$^{\rm c}$]
Single-lined spectroscopic binaries according to the SIMBAD database
are designated with SB1.
\end{list}

\begin{list}{}{}
\item[$^{\rm d}$]
UVES spectrum is also available.
\end{list}

\end{table*}

\begin{table*}
\caption[ ]{Atmospheric parameters, abundance ratios, and space velocities
for stars with VLT/UVES spectra.} 
\label{table:UVES}
\setlength{\tabcolsep}{0.12cm}
\begin{tabular}{lccccrrrrrrrrrrr}
\noalign{\smallskip}
\hline\hline
\noalign{\smallskip}
  ID & \teff  & \logg  & \fracfeh  & \micro  &
  \fracnafe  & \fracmgfe  & \fracsife  & \fraccafe  & \fractife  & \fraccrfe  & \fracnife & $U_{\rm LSR}$ & $V_{\rm LSR}$ & $W_{\rm LSR}$ & Class$^{\rm a}$   \\
     & (K)    &        &       &(km/s)    &
         &        &        &        &        &        &       & (km/s)    &(km/s)     &(km/s)      \\
\hline
\noalign{\smallskip}
BD$-$21 3420  &   5808 &  4.26 & $-$1.13 &   1.3 & $-$0.01 &  0.32 &  0.36 &  0.30 &  0.24 & $-$0.03 & $-$0.01
 &   $-$19 &   $-$87 &   $-$92 &  TD       \\
CD$-$33 3337  &   5979 &  3.86 & $-$1.36 &   1.7 &  0.13 &  0.32 &  0.28 &  0.30 &  0.30 & $-$0.02 &  0.00
 &    14 &   $-$42 &  $-$109 &  TD       \\
CD$-$43 6810  &   5945 &  4.26 & $-$0.43 &   1.3 &  0.15 &  0.28 &  0.23 &  0.21 &  0.19 &  0.03 &  0.01
 &   102 &  $-$159 &   $-$89 &  high-$\alpha$    \\
CD$-$45 3283  &   5597 &  4.55 & $-$0.91 &   1.0 & $-$0.33 &  0.08 &  0.13 &  0.19 &  0.08 & $-$0.01 & $-$0.14
 &  $-$340 &  $-$213 &   $-$99 &   low-$\alpha$    \\
CD$-$51 4628  &   6153 &  4.31 & $-$1.30 &   1.4 & $-$0.26 &  0.13 &  0.19 &  0.31 &  0.24 &  0.00 & $-$0.10
 &   266 &  $-$129 &   175 &   low-$\alpha$    \\
CD$-$57 1633  &   5873 &  4.28 & $-$0.90 &   1.2 & $-$0.33 &  0.03 &  0.06 &  0.16 &  0.03 & $-$0.04 & $-$0.19
 &  $-$329 &  $-$240 &   $-$28 &   low-$\alpha$    \\
CD$-$61 282   &   5759 &  4.31 & $-$1.23 &   1.3 & $-$0.20 &  0.21 &  0.20 &  0.28 &  0.17 & $-$0.02 & $-$0.10
 &   293 &  $-$284 &     7 &   low-$\alpha$    \\
G05-19      &   5854 &  4.26 & $-$1.18 &   1.3 & $-$0.25 &  0.18 &  0.17 &  0.26 &  0.14 & $-$0.05 & $-$0.13
 &   302 &  $-$274 &   $-$89 &   low-$\alpha$    \\
G05-40      &   5795 &  4.17 & $-$0.81 &   1.2 &  0.12 &  0.35 &  0.31 &  0.31 &  0.25 &  0.03 &  0.00
 &   137 &  $-$218 &   $-$86 &  high-$\alpha$    \\
G18-28      &   5372 &  4.41 & $-$0.83 &   1.0 &  0.06 &  0.36 &  0.30 &  0.29 &  0.27 &  0.05 &  0.00
 &   $-$29 &  $-$214 &    32 &  high-$\alpha$    \\
G18-39      &   6040 &  4.21 & $-$1.39 &   1.5 & $-$0.06 &  0.35 &  0.29 &  0.39 &  0.31 & $-$0.01 & $-$0.05
 &  $-$193 &  $-$257 &   $-$19 &  high-$\alpha$    \\
G20-15      &   6027 &  4.32 & $-$1.49 &   1.6 & $-$0.18 &  0.22 &  0.23 &  0.29 &  0.24 & $-$0.03 & $-$0.05
 &   161 &   $-$60 &  $-$210 &  (low-$\alpha$)   \\
G46-31      &   5901 &  4.23 & $-$0.83 &   1.4 & $-$0.13 &  0.15 &  0.13 &  0.20 &  0.11 & $-$0.03 & $-$0.12
 &    23 &  $-$310 &    56 &   low-$\alpha$    \\
G63-26      &   6043 &  4.18 & $-$1.56 &   1.8 & $-$0.08 &  0.35 &  0.49 &  0.37 &  0.28 & $-$0.01 &  0.06
 &    90 &  $-$332 &    43 & (high-$\alpha$)   \\
G66-22      &   5236 &  4.41 & $-$0.86 &   0.9 & $-$0.38 &  0.08 &  0.09 &  0.22 &  0.10 &  0.03 & $-$0.14
 &  $-$268 &  $-$191 &    16 &   low-$\alpha$    \\
G82-05      &   5277 &  4.45 & $-$0.75 &   0.9 & $-$0.41 &  0.06 &  0.05 &  0.18 &  0.08 &  0.02 & $-$0.16
 &  $-$357 &  $-$180 &    53 &   low-$\alpha$    \\
G112-43     &   6074 &  4.03 & $-$1.25 &   1.3 & $-$0.11 &  0.21 &  0.15 &  0.29 &  0.29 &  0.00 & $-$0.02
 &   145 &  $-$119 &  $-$293 &   low-$\alpha$    \\
G112-44     &   5819 &  4.25 & $-$1.29 &   1.2 & $-$0.12 &  0.22 &  0.20 &  0.26 &  0.22 & $-$0.01 & $-$0.02
 &   138 &  $-$109 &  $-$293 &   low-$\alpha$    \\
G114-42     &   5643 &  4.38 & $-$1.10 &   1.3 & $-$0.18 &  0.18 &  0.22 &  0.24 &  0.13 & $-$0.02 & $-$0.09
 &   377 &  $-$189 &   $-$80 &   low-$\alpha$    \\
G121-12     &   5928 &  4.23 & $-$0.93 &   1.4 & $-$0.30 &  0.03 &  0.10 &  0.19 &  0.08 & $-$0.04 & $-$0.19
 &  $-$350 &  $-$202 &    99 &   low-$\alpha$    \\
G188-22     &   5974 &  4.18 & $-$1.32 &   1.5 & $-$0.04 &  0.39 &  0.37 &  0.37 &  0.28 & $-$0.04 & $-$0.01
 &   193 &   $-$99 &    71 &  high-$\alpha$    \\
G159-50     &   5624 &  4.37 & $-$0.93 &   1.1 &  0.02 &  0.38 &  0.31 &  0.32 &  0.24 &  0.02 &  0.00
 &  $-$202 &  $-$192 &    67 &  high-$\alpha$    \\
HD3567      &   6051 &  4.02 & $-$1.16 &   1.5 & $-$0.24 &  0.14 &  0.17 &  0.31 &  0.22 & $-$0.05 & $-$0.13
 &   164 &  $-$259 &   $-$51 &   low-$\alpha$    \\
HD17820     &   5773 &  4.22 & $-$0.67 &   1.3 &  0.13 &  0.35 &  0.30 &  0.27 &  0.25 &  0.02 &  0.03
 &    42 &   $-$85 &   $-$70 &  TD       \\
HD22879     &   5759 &  4.25 & $-$0.85 &   1.3 &  0.10 &  0.36 &  0.32 &  0.30 &  0.25 &  0.00 &  0.00
 &  $-$104 &   $-$76 &   $-$37 &  TD       \\
HD25704     &   5868 &  4.26 & $-$0.85 &   1.4 &  0.08 &  0.32 &  0.25 &  0.22 &  0.19 & $-$0.01 &  0.00
 &  $-$119 &   $-$53 &    $-$2 &  TD       \\
HD51754     &   5767 &  4.29 & $-$0.58 &   1.4 &  0.11 &  0.33 &  0.25 &  0.24 &  0.21 &  0.02 &  0.03
 &   232 &  $-$150 &    13 &  high-$\alpha$    \\
HD59392     &   6012 &  3.91 & $-$1.60 &   1.9 & $-$0.15 &  0.28 &  0.27 &  0.38 &  0.33 &  0.00 & $-$0.01
 &   115 &  $-$307 &   $-$25 &  (low-$\alpha$)   \\
HD76932     &   5877 &  4.13 & $-$0.87 &   1.4 &  0.10 &  0.34 &  0.28 &  0.30 &  0.25 &  0.01 &  0.00
 &   $-$41 &   $-$77 &    76 &  TD       \\
HD97320     &   6008 &  4.19 & $-$1.17 &   1.6 &  0.10 &  0.34 &  0.28 &  0.26 &  0.23 & $-$0.04 &  0.00
 &    80 &   $-$11 &   $-$30 &  TD       \\
HD103723    &   5938 &  4.19 & $-$0.80 &   1.2 & $-$0.17 &  0.10 &  0.11 &  0.21 &  0.12 & $-$0.04 & $-$0.14
 &   $-$72 &  $-$193 &    58 &   low-$\alpha$    \\
HD105004    &   5754 &  4.30 & $-$0.82 &   1.2 & $-$0.05 &  0.17 &  0.14 &  0.17 &  0.07 & $-$0.03 & $-$0.06
 &   $-$44 &  $-$239 &   $-$92 &   low-$\alpha$    \\
HD106516    &   6196 &  4.42 & $-$0.68 &   1.3 &  0.16 &  0.34 &  0.30 &  0.26 &  0.25 &  0.00 &  0.02
 &    62 &   $-$61 &   $-$50 &  TD       \\
HD111980    &   5778 &  3.96 & $-$1.08 &   1.5 &  0.03 &  0.36 &  0.40 &  0.34 &  0.25 & $-$0.02 &  0.00
 &   239 &  $-$174 &   $-$57 &  high-$\alpha$    \\
HD113679    &   5672 &  3.99 & $-$0.65 &   1.4 &  0.15 &  0.37 &  0.33 &  0.32 &  0.26 &  0.02 &  0.03
 &   $-$96 &  $-$278 &     3 &  high-$\alpha$    \\
HD114762A   &   5856 &  4.21 & $-$0.70 &   1.5 &  0.14 &  0.29 &  0.25 &  0.22 &  0.19 &  0.01 &  0.00
 &   $-$71 &   $-$53 &    64 &  TD       \\
HD120559    &   5412 &  4.50 & $-$0.89 &   1.1 &  0.08 &  0.36 &  0.28 &  0.27 &  0.28 &  0.05 &  0.03
 &   $-$19 &   $-$38 &   $-$30 &  TD       \\
HD121004    &   5669 &  4.37 & $-$0.70 &   1.3 &  0.10 &  0.36 &  0.31 &  0.32 &  0.28 &  0.02 &  0.01
 &    70 &  $-$242 &   105 &  high-$\alpha$    \\
HD126681    &   5507 &  4.45 & $-$1.17 &   1.2 & $-$0.06 &  0.38 &  0.42 &  0.34 &  0.25 &  0.02 & $-$0.01
 &   $-$15 &   $-$28 &   $-$64 &  TD       \\
HD132475    &   5646 &  3.76 & $-$1.49 &   1.5 & $-$0.02 &  0.41 &  0.51 &  0.37 &  0.24 & $-$0.03 &  0.01
 &    44 &  $-$371 &    60 & (high-$\alpha$)   \\
HD148816    &   5823 &  4.13 & $-$0.73 &   1.4 &  0.14 &  0.32 &  0.29 &  0.25 &  0.22 &  0.00 &  0.03
 &    98 &  $-$260 &   $-$73 &  high-$\alpha$    \\
HD163810    &   5501 &  4.56 & $-$1.20 &   1.3 & $-$0.22 &  0.18 &  0.17 &  0.28 &  0.22 &  0.02 & $-$0.10
 &   254 &  $-$193 &    37 &   low-$\alpha$    \\
HD175179    &   5713 &  4.33 & $-$0.65 &   1.2 &  0.13 &  0.34 &  0.28 &  0.28 &  0.25 &  0.03 &  0.01
 &    96 &  $-$102 &   $-$19 &  TD       \\
HD179626    &   5850 &  4.13 & $-$1.04 &   1.6 &  0.07 &  0.35 &  0.30 &  0.32 &  0.26 &  0.01 & $-$0.04
 &    91 &  $-$215 &    45 &  high-$\alpha$    \\
HD189558    &   5617 &  3.80 & $-$1.12 &   1.4 & $-$0.04 &  0.36 &  0.36 &  0.35 &  0.26 & $-$0.02 &  0.00
 &    80 &  $-$109 &    48 &  TD       \\
HD193901    &   5656 &  4.36 & $-$1.09 &   1.3 & $-$0.27 &  0.13 &  0.18 &  0.23 &  0.11 & $-$0.04 & $-$0.14
 &  $-$148 &  $-$233 &   $-$66 &   low-$\alpha$    \\
HD194598    &   5942 &  4.33 & $-$1.09 &   1.4 & $-$0.12 &  0.18 &  0.18 &  0.22 &  0.13 & $-$0.04 & $-$0.09
 &   $-$65 &  $-$267 &   $-$29 &   low-$\alpha$    \\
HD199289    &   5810 &  4.28 & $-$1.04 &   1.3 &  0.08 &  0.36 &  0.31 &  0.27 &  0.24 & $-$0.02 &  0.02
 &   $-$32 &   $-$55 &   $-$19 &  TD       \\
HD205650    &   5698 &  4.32 & $-$1.17 &   1.3 &  0.09 &  0.36 &  0.32 &  0.28 &  0.23 & $-$0.02 &  0.00
 &  $-$114 &   $-$71 &    18 &  TD       \\
HD219617    &   5862 &  4.28 & $-$1.45 &   1.5 & $-$0.17 &  0.26 &  0.35 &  0.31 &  0.20 & $-$0.05 & $-$0.08
 &   228 &  $-$174 &   $-$32 &  (low-$\alpha$)   \\
HD222766    &   5334 &  4.27 & $-$0.67 &   0.8 &  0.06 &  0.35 &  0.28 &  0.31 &  0.28 &  0.07 &  0.01
 &  $-$179 &  $-$173 &     0 &  high-$\alpha$    \\
HD241253    &   5831 &  4.31 & $-$1.10 &   1.3 &  0.04 &  0.32 &  0.31 &  0.30 &  0.24 & $-$0.02 & $-$0.01
 &    18 &   $-$74 &    91 &  TD       \\
HD284248    &   6135 &  4.25 & $-$1.57 &   1.6 & $-$0.24 &  0.22 &  0.23 &  0.32 &  0.29 & $-$0.07 & $-$0.07
 &  $-$349 &  $-$150 &   $-$70 &  (low-$\alpha$)   \\
\noalign{\smallskip}
\hline
\end{tabular}

\begin{list}{}{}
\item[$^{\rm a}$]
Classification as thick disk (TD), low-$\alpha$, or high-$\alpha$.
For halo stars with $\feh < -1.4$, the classification
is uncertain and given in parentheses.
\end{list}

\end{table*}

\begin{table*}
\caption[ ]{Atmospheric parameters, abundance ratios, and space velocities
for stars with NOT/FIES spectra.} 
\label{table:FIES}
\setlength{\tabcolsep}{0.12cm}
\begin{tabular}{lccccrrrrrrrrrrr}
\noalign{\smallskip}
\hline\hline
\noalign{\smallskip}
  ID & \teff  & \logg  & \fracfeh  & \micro  &
  \fracnafe  & \fracmgfe  & \fracsife  & \fraccafe  & \fractife  & \fraccrfe  & \fracnife & $U_{\rm LSR}$ & $V_{\rm LSR}$ & $W_{\rm LSR}$ & Class$^{\rm a}$   \\
     & (K)    &        &       & (km/s)    &
         &        &        &        &        &        &       & (km/s)   & (km/s)   & (km/s)    \\
\hline
\noalign{\smallskip}
G05-36      &   6013 &  4.23 & $-$1.23 &   1.4 & $-$0.01 &  0.33 &  0.37 &  0.38 &  0.34 & $-$0.01 & $-$0.01
 &   $-$41 &  $-$297 &   $-$60 &  high-$\alpha$    \\
G13-38      &   5263 &  4.54 & $-$0.88 &   0.9 &  0.05 &  0.35 &  0.31 &  0.30 &  0.32 &  0.06 &  0.00
 &   109 &  $-$143 &    89 &  high-$\alpha$    \\
G15-23      &   5297 &  4.57 & $-$1.10 &   1.0 & $-$0.02 &  0.40 &  0.33 &  0.32 &  0.33 &  0.06 & $-$0.01
 &    19 &  $-$176 &   $-$57 &  high-$\alpha$    \\
G16-20      &   5625 &  3.64 & $-$1.42 &   1.5 & $-$0.13 &  0.22 &  0.33 &  0.32 &  0.19 & $-$0.07 &  0.01
 &   208 &  $-$143 &    75 &  (low-$\alpha$)   \\
G20-15$^{\rm b}$      &   6072 &  4.36 & $-$1.41 &   1.1 & $-$0.23 &  0.20 &  0.15 &  0.25 &  0.22 & $-$0.03 & $-$0.05
 &   161 &   $-$60 &  $-$210 &  (low-$\alpha$)   \\
G21-22      &   5901 &  4.24 & $-$1.09 &   1.4 & $-$0.26 &  0.09 &  0.07 &  0.17 &  0.03 & $-$0.02 & $-$0.16
 &   247 &  $-$258 &   $-$31 &   low-$\alpha$    \\
G24-13      &   5673 &  4.31 & $-$0.72 &   1.0 &  0.08 &  0.34 &  0.27 &  0.29 &  0.25 &  0.02 &  0.01
 &   164 &   $-$38 &  $-$128 &  high-$\alpha$    \\
G24-25      &   5828 &  3.86 & $-$1.40 &   1.2 & $-$0.03 &  0.34 &  0.41 &  0.32 &  0.31 & $-$0.04 &  0.00
 &  $-$197 &  $-$242 &    43 & (high-$\alpha$)   \\
G31-55      &   5638 &  4.30 & $-$1.10 &   1.4 &  0.07 &  0.28 &  0.33 &  0.29 &  0.25 &  0.05 &  0.01
 &   $-$23 &  $-$184 &   $-$45 &  high-$\alpha$    \\
G49-19      &   5772 &  4.25 & $-$0.55 &   1.2 &  0.12 &  0.30 &  0.27 &  0.28 &  0.23 &  0.04 &  0.02
 &    37 &  $-$236 &    58 &  high-$\alpha$    \\
G53-41      &   5859 &  4.27 & $-$1.20 &   1.3 &  0.23 &  0.24 &  0.24 &  0.31 &  0.14 & $-$0.03 & $-$0.09
 &   $-$31 &  $-$299 &  $-$150 &   low-$\alpha$    \\
G56-30      &   5830 &  4.26 & $-$0.89 &   1.3 & $-$0.19 &  0.10 &  0.12 &  0.17 &  0.06 & $-$0.03 & $-$0.16
 &    32 &  $-$283 &   $-$53 &   low-$\alpha$    \\
G56-36      &   5933 &  4.28 & $-$0.94 &   1.4 & $-$0.06 &  0.19 &  0.19 &  0.25 &  0.17 & $-$0.02 & $-$0.08
 &  $-$105 &  $-$247 &     8 &   low-$\alpha$    \\
G57-07      &   5676 &  4.25 & $-$0.47 &   1.1 &  0.12 &  0.34 &  0.29 &  0.29 &  0.30 &  0.05 &  0.05
 &   $-$73 &  $-$162 &   $-$50 &  high-$\alpha$    \\
G74-32      &   5772 &  4.36 & $-$0.72 &   1.1 &  0.10 &  0.37 &  0.29 &  0.30 &  0.26 &  0.02 &  0.02
 &  $-$143 &  $-$151 &    74 &  high-$\alpha$    \\
G75-31      &   6010 &  4.02 & $-$1.03 &   1.4 & $-$0.19 &  0.12 &  0.17 &  0.31 &  0.19 & $-$0.06 & $-$0.13
 &  $-$259 &  $-$206 &   109 &   low-$\alpha$    \\
G81-02      &   5859 &  4.19 & $-$0.69 &   1.3 &  0.06 &  0.25 &  0.19 &  0.20 &  0.11 & $-$0.01 & $-$0.04
 &  $-$175 &  $-$201 &    18 &  high-$\alpha$    \\
G85-13      &   5628 &  4.38 & $-$0.59 &   1.0 &  0.13 &  0.33 &  0.28 &  0.27 &  0.25 &  0.04 &  0.03
 &  $-$184 &   $-$43 &    35 &  high-$\alpha$    \\
G87-13      &   6085 &  4.13 & $-$1.09 &   1.5 & $-$0.25 &  0.10 &  0.18 &  0.32 &  0.22 & $-$0.04 & $-$0.12
 &  $-$201 &  $-$286 &    15 &   low-$\alpha$    \\
G94-49      &   5373 &  4.50 & $-$0.80 &   1.1 &  0.06 &  0.34 &  0.32 &  0.29 &  0.28 &  0.03 &  0.01
 &   172 &   $-$96 &    $-$2 &  high-$\alpha$    \\
G96-20      &   6293 &  4.41 & $-$0.89 &   1.5 &  0.13 &  0.30 &  0.30 &  0.27 &  0.25 & $-$0.06 &  0.00
 &  $-$128 &   $-$91 &   118 &  high-$\alpha$    \\
G98-53      &   5848 &  4.23 & $-$0.87 &   1.3 & $-$0.07 &  0.20 &  0.17 &  0.25 &  0.12 &  0.01 & $-$0.09
 &  $-$156 &  $-$263 &   $-$96 &   low-$\alpha$    \\
G99-21      &   5487 &  4.39 & $-$0.67 &   0.9 &  0.11 &  0.33 &  0.27 &  0.29 &  0.26 &  0.05 &  0.02
 &   $-$64 &  $-$182 &    22 &  high-$\alpha$    \\
G119-64     &   6181 &  4.18 & $-$1.48 &   1.5 & $-$0.15 &  0.25 &  0.21 &  0.37 &  0.29 & $-$0.01 & $-$0.06
 &   237 &  $-$291 &  $-$120 &  (low-$\alpha$)   \\
G125-13     &   5848 &  4.28 & $-$1.43 &   1.5 & $-$0.17 &  0.30 &  0.28 &  0.30 &  0.20 & $-$0.09 & $-$0.09
 &   215 &  $-$228 &  $-$157 & (high-$\alpha$)   \\
G127-26     &   5791 &  4.14 & $-$0.53 &   1.2 &  0.10 &  0.31 &  0.22 &  0.22 &  0.20 &  0.03 &  0.00
 &   171 &   $-$83 &   $-$50 &  high-$\alpha$    \\
G150-40     &   5968 &  4.09 & $-$0.81 &   1.4 &  0.28 &  0.14 &  0.19 &  0.19 &  0.12 &  0.01 & $-$0.08
 &   $-$63 &  $-$318 &    14 &   low-$\alpha$    \\
G161-73     &   5986 &  4.00 & $-$1.00 &   1.4 & $-$0.36 &  0.13 &  0.13 &  0.25 &  0.14 & $-$0.06 & $-$0.15
 &   228 &  $-$250 &    42 &   low-$\alpha$    \\
G170-56     &   5994 &  4.12 & $-$0.92 &   1.5 & $-$0.11 &  0.17 &  0.14 &  0.24 &  0.12 & $-$0.02 & $-$0.11
 &   $-$52 &  $-$289 &   $-$29 &   low-$\alpha$    \\
G172-61     &   5225 &  4.47 & $-$1.00 &   0.9 & $-$0.18 &  0.16 &  0.17 &  0.26 &  0.16 &  0.04 & $-$0.10
 &   $-$14 &  $-$267 &    57 &   low-$\alpha$    \\
G176-53     &   5523 &  4.48 & $-$1.34 &   1.0 & $-$0.36 &  0.15 &  0.15 &  0.25 &  0.15 & $-$0.01 & $-$0.12
 &  $-$230 &  $-$271 &    61 &   low-$\alpha$    \\
G180-24     &   6004 &  4.21 & $-$1.39 &   1.5 & $-$0.14 &  0.31 &  0.31 &  0.39 &  0.31 &  0.03 & $-$0.04
 &   111 &  $-$245 &   $-$21 &  high-$\alpha$    \\
G187-18     &   5607 &  4.39 & $-$0.67 &   1.1 &  0.06 &  0.29 &  0.26 &  0.26 &  0.23 &  0.02 &  0.00
 &   109 &  $-$130 &    72 &  high-$\alpha$    \\
G192-43     &   6170 &  4.29 & $-$1.34 &   1.5 & $-$0.21 &  0.19 &  0.23 &  0.33 &  0.29 & $-$0.05 & $-$0.09
 &  $-$327 &  $-$224 &   $-$15 &   low-$\alpha$    \\
G232-18     &   5559 &  4.48 & $-$0.93 &   1.3 & $-$0.06 &  0.35 &  0.34 &  0.34 &  0.26 &  0.05 &  0.00
 &    72 &  $-$242 &    99 &  high-$\alpha$    \\
HD148816$^{\rm b}$    &   5840 &  4.14 & $-$0.70 &   1.4 &  0.14 &  0.32 &  0.27 &  0.26 &  0.20 &  0.02 &  0.02
 &    98 &  $-$260 &   $-$73 &  high-$\alpha$    \\
HD159482    &   5737 &  4.31 & $-$0.73 &   1.3 &  0.15 &  0.34 &  0.32 &  0.30 &  0.25 &  0.05 &  0.02
 &  $-$164 &   $-$47 &    95 &  high-$\alpha$    \\
HD160693    &   5714 &  4.27 & $-$0.49 &   1.1 &  0.05 &  0.24 &  0.19 &  0.21 &  0.14 & $-$0.01 & $-$0.01
 &   215 &  $-$105 &    92 &  high-$\alpha$    \\
HD177095    &   5349 &  4.39 & $-$0.74 &   0.9 &  0.05 &  0.38 &  0.30 &  0.31 &  0.28 &  0.02 &  0.01
 &   140 &  $-$138 &   $-$23 &  high-$\alpha$    \\
HD179626$^{\rm b}$    &   5855 &  4.19 & $-$1.00 &   1.4 &  0.06 &  0.36 &  0.32 &  0.34 &  0.26 &  0.03 & $-$0.05
 &    91 &  $-$215 &    45 &  high-$\alpha$    \\
HD189558$^{\rm b}$    &   5623 &  3.81 & $-$1.12 &   1.4 & $-$0.01 &  0.38 &  0.38 &  0.36 &  0.27 & $-$0.01 &  0.01
 &    80 &  $-$109 &    48 &  TD       \\
HD193901$^{\rm b}$    &   5676 &  4.41 & $-$1.07 &   1.4 & $-$0.30 &  0.17 &  0.16 &  0.23 &  0.11 & $-$0.04 & $-$0.14
 &  $-$148 &  $-$233 &   $-$66 &   low-$\alpha$    \\
HD194598$^{\rm b}$    &   5926 &  4.32 & $-$1.08 &   1.4 & $-$0.10 &  0.23 &  0.21 &  0.22 &  0.14 & $-$0.03 & $-$0.09
 &   $-$65 &  $-$267 &   $-$29 &   low-$\alpha$    \\
HD230409    &   5318 &  4.54 & $-$0.85 &   1.1 &  0.05 &  0.30 &  0.25 &  0.26 &  0.26 &  0.06 &  0.00
 &   164 &  $-$114 &    $-$4 &  high-$\alpha$    \\
HD233511    &   6006 &  4.23 & $-$1.55 &   1.3 & $-$0.19 &  0.36 &  0.28 &  0.40 &  0.30 & $-$0.05 & $-$0.05
 &  $-$131 &  $-$243 &    31 & (high-$\alpha$)   \\
HD237822    &   5603 &  4.33 & $-$0.45 &   1.1 &  0.15 &  0.35 &  0.26 &  0.28 &  0.26 &  0.04 &  0.05
 &    81 &  $-$131 &   131 &  high-$\alpha$    \\
HD250792A   &   5489 &  4.47 & $-$1.01 &   1.1 & $-$0.18 &  0.23 &  0.13 &  0.33 &  0.26 &  0.04 & $-$0.12
 &   245 &  $-$237 &   104 &   low-$\alpha$    \\
\noalign{\smallskip}
\hline
\end{tabular}

\begin{list}{}{}
\item[$^{\rm a}$]
Classification as thick disk (TD), low-$\alpha$, or high-$\alpha$.
For halo stars with $\feh < -1.4$, the classification
is uncertain and given in parentheses.
\end{list}

\begin{list}{}{}
\item[$^{\rm b}$]
Data are also available from UVES spectra (see Table \ref{table:UVES}).
In the figures, abundance data based on the UVES spectra are plotted.
\end{list}

\end{table*}

\end{document}